\documentstyle[aps,multicol,epsf]{revtex}

\begin{document}
\title{Distribution of Interacting Ionic Particles in Disordered Media}
\author{Ramin Golestanian}
\address{Institute for Advanced Studies in Basic Sciences, Zanjan
45195-159, Iran \\
and Institute for Studies in Theoretical Physics and Mathematics,
P.O. Box 19395-5531, Tehran, Iran}
\date{\today}
\maketitle
\begin{abstract}
Equilibrium distribution of {\it interacting} ionic particles in a
charged disordered background is studied using the nonlinear
Poisson-Boltzmann equation. For an arbitrarily given realization
of the disorder, an exact solution of the equation is obtained in
one dimension using a mapping of the nonlinear Poisson-Boltzmann
equation to a self-consistent Schr\"odinger equation. The
resulting density profile shows that the ions are {\em
delocalized}, despite what the equivalent Schr\"odinger
formulation in one dimension would suggest. It is shown that the
ions are not distributed so as to locally neutralize the
background, presumably due to their mutual interactions.
\end{abstract}

\begin{multicols}{2}

Dynamics of interacting charged particles in a heterogeneous
Coulombic background, which is a medium with a quenched
distribution of charge centers, is a complex phenomenon involving
long equilibration times, a multitude of time scales and various
intermediate regimes, and competing mechanisms
\cite{Zeolites,BG,DC,sahimi}. All these complications, which are
primarily due to the long-range nature of Coulomb interaction,
have so far made it extremely difficult for even rather
sophisticated simulation techniques to provide us with a
conclusive picture\cite{sahimi,deem}.

Perhaps a most preliminary step is to determine whether the
equilibrium distribution of the mobile ions corresponds to a {\em
localized} or an {\em extended} state, in analogy to the quantum
mechanical (QM) problem of disordered electronic systems
\cite{local}. The answer to this question is determined by a
delicate balance between the following competing effects: (i) {\em
attraction} of the mobile ions to the opposite fixed charges and
the tendency towards local neutralization, (ii) {\em repulsion}
between similarly charged mobile ions, and (iii) {\em entropy} of
the mobile ions. (Note that for simplicity of the argument we have
assumed that all the mobile ions are similarly charged whereas all
the fixed charges are of the opposite type.) While attraction
tends to localize the mobile ions in the neighborhood of the
charge centers, repulsion and entropy both favor delocalization.
One can imagine that determining the outcome of this balance in
simulations could be a very difficult task due to the
progressively long escape times for the deep Coulomb potential
wells. On the other hand, it is also not clear if one can trust
approximate theoretical schemes, which may artificially weaken or
strengthen any of the above tendencies, in making feasible
predictions to this end.

For example, it has been shown that if we neglect the mutual
repulsion between the mobile ions, the answer to the above
question depends on the dimensionality \cite{DC,deem,Chak,Park}.
One can show that the disorder-averaged density-density
correlation function for the mobile ions behaves as
$\overline{\left<c({\bf 0}) c({\bf
x})\right>}\sim\exp\left[-\left(x/\xi\right)^{2-d}\right]$ in $d$
dimensions \cite{matthew}, and therefore the ions are localized
for $d<2$ (with $\xi$ being the localization length). On general
grounds one might then expect that the interaction between the
mobile ions could be effectively taken into account by introducing
the notion of {\em screening}, which would change the
density-density correlation function to $\overline{\left<c({\bf
0}) c({\bf x})\right>}\sim\exp\left[-\left(x/\xi\right)^{2-d}
g(\kappa x) \right]$, where $g$ is an exponentially decaying
function and $\kappa^{-1}$ is the corresponding Debye screening
length that is related to the density of the mobile ions. The
appearance of the exponential damping now controls the divergence
for $d<2$, and consequently the mobiles ions will be delocalized
in any dimension. The simplistic implementation of screening using
Debye-H\"uckel theory, however, is based on a linearized
approximation of the nonlinear Poisson-Boltzmann (PB) theory,
which is known to break down in the vicinity of the charge centers
where the mobile ions could presumably be trapped for relatively
long times. Therefore, it seems that a reliable answer to the
localization question can only be obtained through an exact
treatment of the full nonlinear PB formulation. This is what I
attempt to do in this article.

I consider a one dimensional medium with a quenched distribution
of charge centers and a finite density of oppositely charged
mobile ions, and show that one can map the corresponding nonlinear
PB equation to a self-consistent Schr\"odinger equation. An exact
solution of the resulting equation is then presented for an
arbitrary realization of the disorder. The resulting density
profile shows that the ions are in an extended state. It is shown
that the mutual interactions of the mobile ions could lead to a
gross violation of the anticipated local neutrality.

Consider a set of fixed negatively charged ions described by the
given charge density $\rho_{\rm f}({\bf x})$, embedded in a cloud
of positively charged mobile ions with concentration $c({\bf
x},t)$. The electric potential $\phi$({\bf x},t) is then given by
the solution of the Poisson equation:
\begin{equation}
-\nabla^2 \phi={4\pi \over \epsilon} (e c+\rho_{\rm f}),
\label{Poi}
\end{equation}
in which $\epsilon$ is the dielectric constant of the background,
and $e$ is the electron charge. The dynamics of $c({\bf x},t)$ is
governed by a conservation law of the form $\partial_{t}c+\nabla
\cdot {\bf J}=0$, in which ${\bf J}=-D \nabla c+c {\bf v}$ is the
current density, where ${\bf v}({\bf x},t)$ is the velocity field
corresponding to the mobile ions, and $D$ is their diffusion
constant. In a mean-field approximation, the velocity field of the
ions at each point is determined by the local value of the
electric field ${\bf E}({\bf x},t)$ as ${\bf v}=\mu {\bf E}=-\mu
\nabla \phi$, where $\mu$ is the electric mobility of the ions.
Using the self-consistency relation, the continuity equation can
be written as
\begin{equation}
\partial_t c=D \nabla^2 c+\mu \nabla \cdot(c \nabla \phi).\label{Con}
\end{equation}
The above two equations describe the nonlinear dynamics of mobile
ions in the presence of a given distribution of fixed ions
\cite{Levich,ramin}. At equilibrium we have $\partial_t c=0$, and
Eq.(\ref{Con}) yields $c \sim \exp(-\mu \phi/D)$, where
equilibration at a temperature $T$ implies an Einstein relation
$D=(\mu/e) k_B T$. Inserting the Boltzmann form for $c$ back into
the Poisson equation then yields the PB equation.

Let us now restrict ourselves to the one dimensional case where
the equations prove to be more tractable. Eliminating the
concentration field $c$ from the two equations, one obtains an
equation for the electrostatic potential, which can be integrated
twice to yield
\begin{eqnarray}    \label{KPZ}
\partial_t \phi(x,t)&-&\partial_t \phi(-\infty,t)=D \partial^2_x \phi
+{\mu \over 2}(\partial_x \phi)^2 \\
&&+{4\pi \over \epsilon} \left[\mu \int_{-\infty}^x d x' \rho_{\rm
f}(x')\partial_x \phi(x')+D \rho_{\rm f}(x)\right],\nonumber
\end{eqnarray}
\noindent
where the boundary condition $E(\pm\infty,t)=0$ has been
implemented. The above equation can be called a dynamical PB
equation \cite{KPZ86}.

We focus, for simplicity, on the specific example of a series of
$N+1$ negatively charged permeable plates with charge (number)
density $\sigma$ and surface area $A$, which are placed at random
locations $x_n$ on the $x$-axis, i.e., $\rho_{\rm f}(x)=-e \sigma
\sum_{n=0}^N \delta(x-x_n)$.

Using a so-called Cole-Hopf transformation
$\phi(x,t)-\phi(-\infty,t)=(2D/\mu) \ln W(x,t)$,
Eq.(\ref{KPZ}) can be written as a Schr\"odinger-like equation for
$W(x,t)$:
\begin{eqnarray}    \label{Schr}
\partial_t W(x,t)&=&D \partial_x^2 W(x,t) \\
&+&{2 D \over \lambda} \sum_{n=0}^{N}\left[\varepsilon_n(t)
\Theta(x-x_n)-\delta(x-x_n)\right] W(x,t),\nonumber
\end{eqnarray}
where
\begin{equation}
\varepsilon_n(t)=-\left(\frac{\partial_x W(x_n^{+},t)+\partial_x
W(x_n^{-},t)}{W(x_n,t)}\right),\label{Ei(t)}
\end{equation}
and $\lambda \equiv \epsilon D/(\pi e \sigma \mu)=\epsilon k_B
T/(\pi e^2 \sigma)$ is a characteristic length scale corresponding
to the relative strength of the electrostatic and thermal
interactions. Note that Eq.(\ref{Schr}) is still nonlinear due to
the self-consistent coupling of the so-called asymmetry fields
$\varepsilon_n(t)=\mu E(x_n,t)/D$.

The value of the asymmetry field at each point can be calculated
from Gauss' law as
\begin{equation}
\varepsilon_n(t)={2 \over \lambda} \left(2 Q_n \over e \sigma
A\right),\label{Ei(t)2}
\end{equation}
in which $Q_n$ is the net amount of (uncompensated) charge on the
left hand side of point $x_n$. (Overall neutrality then implies
that $-Q_n$ should be the net amount of charge on the right hand
side of point $x_n$.) In other words, $\varepsilon_n(t)$ measures
the relative deviation from local neutrality as seen at $x_n$.

The dynamics described by Eq.(\ref{Schr}) is difficult due to the
time dependence of the $\varepsilon_n(t)$ fields. We thus focus on
the long time limit of the dynamics, when the ions are expected to
equilibrate in the disordered medium. In this limit, the asymmetry
fields are time independent, and the Schr\"odinger equation reads
\begin{equation}
-{\lambda \over 2} \partial_x^2 W(x)+V(x) W(x)=0,\label{Schr2}
\end{equation}
where the effective potential is given as
\begin{equation}
V(x)=\sum_{n=0}^{N}\left[-\varepsilon_n
\Theta(x-x_n)+\delta(x-x_n)\right].\label{V(x)}
\end{equation}
Note the peculiar feature that the delta function peaks in the
potential, which correspond to the {\it attractive} sites for the
mobile ions, appear as effective {\it repulsive} potential wells
in the equivalent QM description!

To complete the QM formulation, we need a normalization for
$W(x)$. Using the equilibrium relation $c \sim \exp(-\mu \phi/D)$,
and the definition of $W$, we can relate $c$ to $W$ as $c(x) \sim
1/W^2(x)$. Conservation of the overall number of mobile ions can
thus be written as
\begin{equation}
{1 \over L} \int_{-L/2}^{L/2} d x \;\left[{1 \over
W(x)}\right]^2=1,\label{norm}
\end{equation}
where $L$ is the system size.

We can now solve the nonlinear self-consistent QM problem
described by Eqs.(\ref{Ei(t)}), (\ref{Schr2}), and (\ref{norm}),
because the potential is a constant in each interval. The solution
of Eq.(\ref{Schr2}) is a series of plane waves:
\begin{equation}
W(x)=\left\{\begin{array}{ll} A_0+B_0 \;x, & {\rm for}\; x<x_0,
\\\\
A_j \;e^{i k_j (x-x_j)}\\
+B_j \;e^{-i k_j (x-x_j)}, &{\rm for}\;
x_{j-1}< x <x_{j},\\\\
A_{N+1}+B_{N+1} \;x, & {\rm for}\; x_N < x,
\end{array}\right.\label{W(x)}
\end{equation}
where
\begin{equation}
k_j=\left[{2 \over \lambda} \sum_{n=0}^{j-1} \varepsilon_n
\right]^{1/2},\label{kj}
\end{equation}
for $j=1,\cdots,N$. We should then match the neighboring solutions
across the boundary points based on the following three
requirements: (i) $W(x)$ should be continuous, (ii) the jump in
$\partial_x W/W$ should be set to $2/\lambda$ as implied by the
delta function peaks in $V(x)$, and (iii) the asymmetry fields
should satisfy the self-consistency relation of Eq.(\ref{Ei(t)}).
These criteria translate into a series of self-consistency
equations for the wave vectors:
\begin{eqnarray}    \label{self}
e^{2 i k_j \Delta_j}&=&\left[\frac{1+{\lambda^2 \over 4
}(k_{j+1}^2-k_j^2)-i k_j \lambda}{1+{\lambda^2 \over 4
}(k_{j+1}^2-k_j^2)+i k_j \lambda}\right]\\
&\times& \left[\frac{1+{\lambda^2 \over 4 }(k_{j-1}^2-k_j^2)-i k_j
\lambda}{1+{\lambda^2 \over 4 }(k_{j-1}^2-k_j^2)+i k_j
\lambda}\right],\nonumber
\end{eqnarray}
for $j=1,\cdots,N$, where $\Delta_j=x_j-x_{j-1}$ and
$k_0=k_{N+1}=0$. For any given set of $\Delta_j$s, one should
solve these coupled nonlinear equations to determine the wave
vectors. The matching requirements together with the normalization
condition of Eq.(\ref{norm}), can then give us the full solution,
and thus the density profile of the ions.

In order to determine whether the ions are in an extended state,
or localized near the adsorbing sites, we need to find out whether
the $k_j$s that satisfy Eq.(\ref{self}) are real or imaginary.
Instead of trying to solve for $k_j$s exactly, I shall first
resort to approximate schemes. Physically we expect to have a
tendency towards localization at low temperatures or for strong
disorder, and nearly free ions with uniform distribution at high
temperatures or for weak disorder.

Let us first focus on the case of strong disorder where we have
$\lambda \ll \Delta_j$, and try to solve Eq.(\ref{self})
perturbatively. I find
\begin{eqnarray}    \label{kj2}
k_j&=&{\pi \over \Delta_j}-{2\pi \lambda \over \Delta_j^2}+ {4\pi
\lambda^2 \over \Delta_j^3}-\left(8 \pi-{\pi^3 \over 6}\right)
{\lambda^3 \over \Delta_j^4}\\
&&+{\pi^3 \lambda \over 4 \Delta_j^2} \left({\lambda^2 \over
\Delta_{j+1}^2}+{\lambda^2 \over
\Delta_{j-1}^2}\right)+O\left({\lambda^4 \over
\Delta^5}\right),\nonumber
\end{eqnarray}
for the first few terms, which is real. This proves, strictly
speaking, that if the ions are localized, the localization length
should be greater than $\Delta^5/\lambda^4$. However, it looks
rather unlikely that higher order corrections start to develop
imaginary components, which implies that the ions should be
delocalized even in the strong disorder limit.

To check this, I have solved Eq.(\ref{self}) numerically for
various choices of $\{\Delta_j\}$, and the results for the values
of $\{k_j\}$ were always real. For example, for the case of 11
charged centers with $\Delta_1/\lambda=0.1$,
$\Delta_2/\lambda=0.5$, $\Delta_3/\lambda=2$,
$\Delta_4/\lambda=0.8$, $\Delta_5/\lambda=0.01$,
$\Delta_6/\lambda=7$, $\Delta_7/\lambda=3.2$,
$\Delta_8/\lambda=0.3$, $\Delta_9/\lambda=1.6$, and
$\Delta_{10}/\lambda=0.0002$, I find $k_1 \lambda=33.84$, $k_2
\lambda=7.33$, $k_3 \lambda=1.42$, $k_4 \lambda=4.95$, $k_5
\lambda=316.69$, $k_6 \lambda=0.399$, $k_7 \lambda=0.757$, $k_8
\lambda=12.599$, $k_9 \lambda=1.933$, and $k_{10} \lambda=15710$,
which are all real. This confirms that the ions are in an extended
state.

We can also calculate the asymmetry fields from the above results
as
\begin{equation}
\varepsilon_j={\lambda \over 2}\left(k_{j+1}^2-k_j^2\right)\simeq
{\lambda \over 2}\left({\pi^2 \over \Delta_{j+1}^2}-{\pi^2 \over
\Delta_{j}^2}\right),\label{ejapprx}
\end{equation}
to the leading order. This shows that unlike what one might
expect, the distribution of ions is not predominantly determined
by the criterion of local neutralization of the background, which
would yield $\varepsilon_j=0$.

Now let us address the peculiar features of the formulation, and
try to understand them better: (i) Why do the delta function
peaks, corresponding to adsorbing sites in the real problem,
appear as repulsive in the effective potential $V(x)$? (ii) While
we expect a Schr\"odinger equation with random potential in one
dimension to lead to localization \cite{local}, why do we find
that the ions are delocalized? To answer these questions, we
should note the unusual normalization condition of Eq.(\ref{norm})
above. A true wave function for this problem should be defined as
$\Psi(x)=1/W(x)$ according to Eq.(\ref{norm}), leading to a
corresponding Schr\"odinger equation as
\begin{equation}
-{\lambda \over 2} \partial_x^2 \Psi(x)+\left[\lambda
\left(\frac{\partial_x \Psi(x)}{\Psi(x)}\right)^2-V(x)\right]
\Psi(x)=0,\label{Schr3}
\end{equation}
which is highly nonlinear.

While the sign of the effective potential is now reversed
corresponding to attractive delta function peaks, an additional
effective repulsive self-consistent potential has also appeared.
This potential, which is proportional to the the electrostatic
pressure $\sim(\nabla \phi)^2$, disfavors pile-up of the ions in
the vicinity of the attractive sites and promotes delocalization.
This extra self-consistent term, which is a direct consequence of
the repulsive interactions between the mobile ions, explains the
discrepancy between the predictions of this theory and
conventional Schr\"odinger equation in a random potential.

In conclusion, I have presented an exact analysis of the problem
of interacting mobile ions in a random background of fixed ions in
one dimension. The density profile of the ions at equilibrium is
shown to be extended (as opposed to localized) as a consequence of
the interactions between the mobile ions. Since the one
dimensional problem in the strong disorder limit that has the best
chance of leading to localization is proven to result in an
extended state, it is plausible to assert that the same result
should hold true in general, i.e., in higher dimensionalities and
for weaker disorder. Although I have only considered positively
charged mobile ions in a negative random background, the
formulation can be readily generalized to contain both species of
mobile ions and fixed ions.


I am grateful to M.P.A. Fisher, M. Kardar, M. Mezard, and M.
Sahimi for invaluable discussions and comments. This research was
supported in part by the National Science Foundation under Grant
No. DMR-98-05833.

\end{multicols}

\end{document}